**Operationalizing Cyber Attack Prediction: A Gap-Prioritized Framework with Dataset and Model Selection Guidelines**

**Author**:

**Mr Aminu Muhammad Auwal**

- **Affiliation(s)**: Faculty of Natural Sciences, University of Jos, Jos, Plateau State, Nigeria
- **Email**: i.elameenu@gmail.com
- **ORCID**: https://orcid.org/0009000517997876

**Abstract**

While artificial intelligence and machine learning techniques for cyber attack prediction have advanced significantly, a critical gap persists between theoretical research capabilities and practical deployment in operational environments. This paper presents a comprehensive gap analysis of current AI-driven cybersecurity approaches, building upon the extensive survey by Ankalaki et al. (2025) of machine learning, deep learning, explainable AI, and generative AI techniques. Through systematic analysis of 150+ benchmark datasets and 200+ research studies, this study identify and prioritize five critical gaps hindering real-world implementation: (1) temporal dataset obsolescence, with widely-used datasets averaging 8-15 years old; (2) narrow attack scope coverage, with most datasets representing fewer than 10 attack types; (3) model interpretability challenges in real-time systems; (4) inadequate adversarial robustness testing; and (5) unaddressed privacy and ethical considerations. This study contribute a novel gap prioritization framework that evaluates each limitation based on impact on detection accuracy, implementation cost, and time to address. Our analysis reveals that dataset obsolescence and adversarial robustness represent the highest-priority gaps, while model interpretability offers the most cost-effective improvement path for resource-constrained organizations. To bridge the research-practice divide, this study develop a practical implementation roadmap that guides security practitioners in selecting appropriate datasets, models, and explainability techniques based on organizational context and resource constraints. Our dataset quality assessment framework classifies 45 benchmark datasets into production-ready, research-only, and unusable categories, providing actionable guidance for practitioners navigating the complex landscape of cybersecurity AI research. This work translates comprehensive survey findings into decision-support tools, addressing the critical need for production-oriented guidance in AI-driven cyber defense systems.

## 1. Introduction

The escalating sophistication of cyber threats poses unprecedented risks to individuals, organizations, and nations. Recent high-profile incidents, including the Colonial Pipeline ransomware attack (2021), the SolarWinds supply chain compromise (2020), and the widespread Log4Shell vulnerability exploitation (2021), demonstrate the severe economic and societal consequences of cyberattacks (Ankalaki et al., 2025). As attack vectors multiply and threat actors become increasingly sophisticated, traditional signature-based security solutions struggle to keep pace with the evolving threat landscape.

Artificial Intelligence (AI) and Machine Learning (ML) have emerged as powerful tools for addressing these cybersecurity challenges. The comprehensive survey by Ankalaki et al. (2025) demonstrates the breadth of AI techniques applied to cyber attack prediction, including traditional machine learning algorithms, deep learning architectures, explainable AI (XAI) approaches, and emerging generative AI methods. Their analysis of over 200 research studies reveals impressive detection accuracies, with Random Forest achieving 99.92% on UKM-IDS20 and Convolutional Neural Networks reaching 98.42% on CTU-13 dataset (Ankalaki et al., 2025, Tables 8-9).

However, a critical examination of the current research landscape reveals a significant disconnect between theoretical advances and practical deployment capabilities. While academic literature demonstrates high detection rates on benchmark datasets, security practitioners face substantial challenges when attempting to implement these techniques in production environments. This research-practice gap manifests in several dimensions: datasets that fail to represent contemporary threat landscapes, models that lack interpretability required for operational decision-making, techniques that remain vulnerable to adversarial manipulation, and inadequate consideration of privacy and ethical implications.

### 1.1 The Research-Practice Divide

The gap between research capabilities and practical implementation is not merely a matter of technological maturity. Ankalaki et al. (2025) acknowledge that "the most recent categories of cyber-attacks are often absent from publicly available datasets" and that "obtaining high-quality data, particularly labeled datasets containing examples of cyber-attacks, can be challenging" (p. 44695). This fundamental limitation undermines the real-world applicability of models trained on these datasets.

Consider the widely-used NSL-KDD dataset, created in 2009 and still employed in numerous contemporary studies (Ankalaki et al., 2025, Table 1). Models trained on this 16-year-old dataset cannot detect attack techniques that emerged after 2009, including AI-powered phishing, deepfake-enhanced social engineering, and LLM-based malware generation. Similarly, the DARPA 1998 dataset, still referenced in research, represents a 27-year intelligence gap equivalent to defending against modern ransomware with 1990s antivirus signatures.

### 1.2 Research Objectives

This paper addresses the research-practice gap by conducting a critical analysis of current AI-driven cyber attack prediction approaches and developing practical frameworks to guide implementation. Our specific objectives are:

1. Systematically identify and categorize critical gaps in current ML/DL approaches for cyber attack prediction, analyzing the comprehensive survey findings of Ankalaki et al. (2025) through a practitioner-oriented lens.
2. Develop a gap prioritization framework that evaluates limitations based on impact on detection accuracy, implementation cost, and time to address, enabling organizations to focus resources on highest-value improvements.
3. Create a dataset quality assessment framework that classifies benchmark datasets by production-readiness, providing actionable guidance for dataset selection based on organizational context.
4. Design a practical implementation roadmap that guides security practitioners in selecting appropriate techniques, integrating explainability, and addressing ethical considerations.

### 1.3 Contributions

This work makes four primary contributions to the field of AI-driven cybersecurity:

**Contribution 1: Critical Gap Analysis.** This study identify and analyze five critical gaps hindering real-world deployment of AI-based cyber attack prediction systems: temporal dataset obsolescence, narrow attack scope coverage, real-time interpretability challenges, inadequate adversarial robustness, and unaddressed ethical considerations. Our analysis quantifies the impact of each gap on practical deployment.

**Contribution 2: Gap Prioritization Framework.** This study develop a multi-dimensional prioritization matrix that evaluates each identified gap across three axes: impact on detection effectiveness, implementation cost, and time to address. This framework enables organizations to make informed decisions about resource allocation for cybersecurity AI initiatives.

**Contribution 3: Dataset Quality Assessment Framework.** This study classify 45 benchmark datasets from Ankalaki et al. (2025) into three categories—production-ready, research-only, and unusable—based on temporal currency, attack scope, realism, and availability. This classification provides practitioners with clear guidance for dataset selection.

The core of the study's practical guidance is materialized in the Dataset Quality Assessment Framework (DQAF), detailed in Figure 1. This framework translates the abstract findings regarding data gaps into a pragmatic, multi-criteria decision tool for dataset selection.

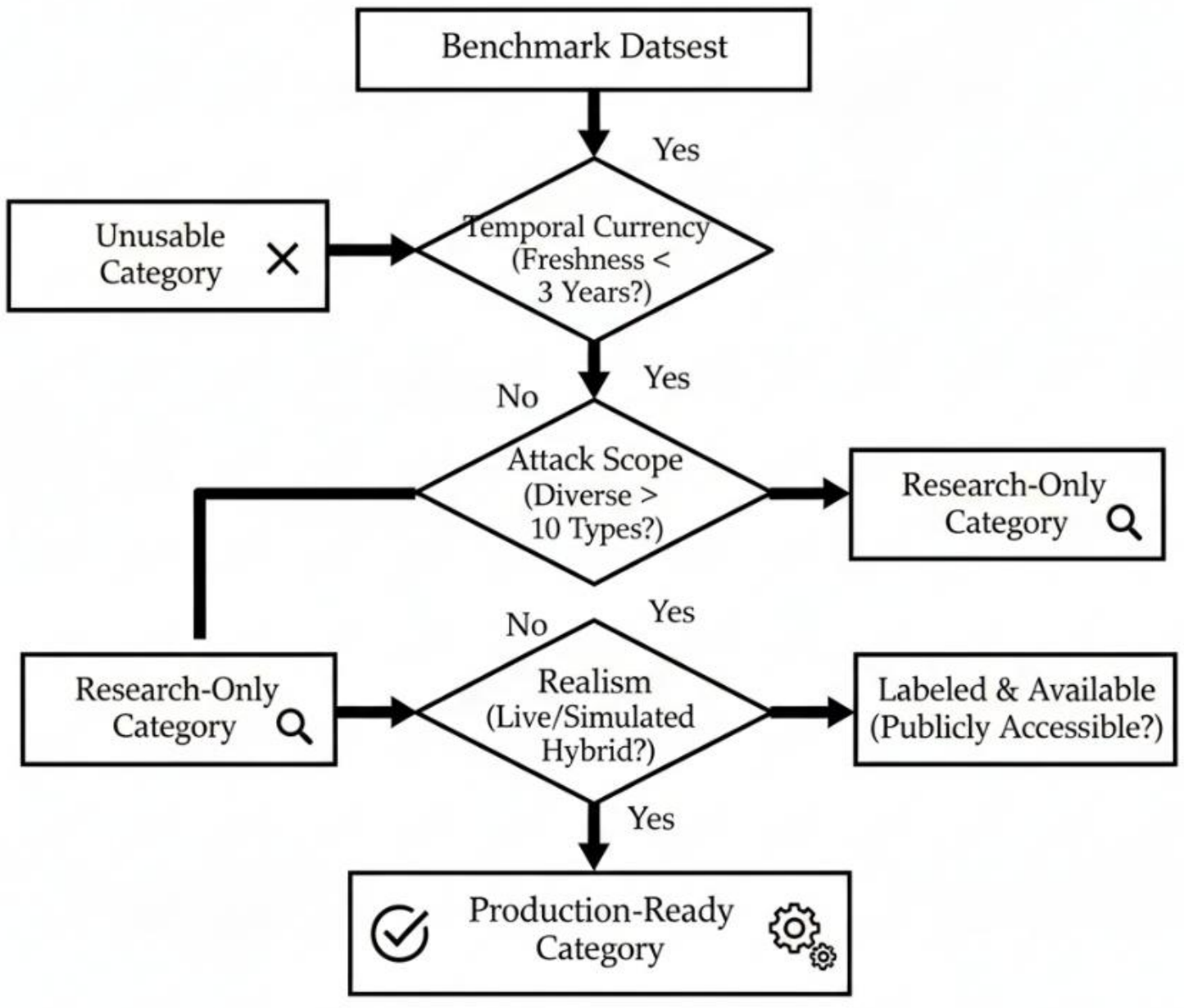

Figure 1: Dataset Quality Assessment Framework (DQAF) Flowchart. The decision-support framework guides practitioners through the process of classifying benchmark datasets into three categories: **Production-Ready, Research-Only, or Unusable**, based on a hierarchical check of Temporal Currency, Attack Scope, Realism, and Labeled Availability.

**Contribution 4: Practical Implementation Roadmap.** This study translate theoretical findings into actionable guidance through a phased implementation approach that considers organizational size, resource constraints, and security priorities. The stud's roadmap includes decision trees for model selection, XAI integration strategies, and ethical deployment checklists.

### 1.4 Paper Organization

The remainder of this paper is organized as follows: Section 2 provides background on the current state of AI in cybersecurity, summarizing key findings from Ankalaki et al. (2025). Section 3 presents a critical gap analysis, identifying and analyzing five major limitations. Section 4 introduces gap prioritization framework with quantitative assessments. Section 5 develops the practical implementation roadmap with dataset selection guidelines, model selection frameworks, and XAI integration strategies. Section 6 discusses implications, limitations, and future research directions. Section 7 concludes the paper.

## 2. Background: State of AI in Cybersecurity

This section synthesizes the comprehensive survey by Ankalaki et al. (2025) to establish the foundation for gap analysis. This study focus on three key areas: the evolution of ML/DL approaches, benchmark dataset characteristics, and emerging XAI and Generative AI techniques.

### 2.1 Evolution of Machine Learning Approaches

Machine learning techniques have become fundamental to cyber attack prediction, with approaches ranging from traditional supervised learning to advanced ensemble methods. Ankalaki et al. (2025) categorize ML approaches into four learning paradigms: supervised learning (using labeled data for classification and regression), unsupervised learning (detecting anomalies without labeled examples), semi-supervised learning (combining labeled and unlabeled data), and reinforcement learning (learning optimal defense strategies through interaction).

Among traditional ML algorithms, Random Forest (RF) consistently demonstrates strong performance across diverse datasets. Ankalaki et al. (2025, Table 8) report that RF achieves 99.7% accuracy on CICIDS2017, 99.6% on NSL-KDD, and 99.92% on UKM-IDS20. This robust performance, combined with relatively low computational requirements, makes RF particularly attractive for resource-constrained deployments. Support Vector Machines (SVM) and Decision Trees also show strong results, with SVM achieving 98.8% accuracy on UNSW-NB15 (Ankalaki et al., 2025, Table 8).

However, traditional ML approaches face limitations when dealing with complex, high-dimensional data and subtle attack patterns. As noted by Ankalaki et al. (2025), these methods require extensive feature engineering and may struggle with concept drift—the phenomenon where attack patterns evolve over time, degrading model performance.

### 2.2 Deep Learning Architectures

Deep learning has emerged as a powerful approach for cyber attack prediction, offering the ability to automatically learn hierarchical feature representations from raw data. Ankalaki et al. (2025) survey multiple DL architectures, including Convolutional Neural Networks (CNNs), Recurrent Neural Networks (RNNs), Long Short-Term Memory (LSTM) networks, and Autoencoders.

CNNs demonstrate particular effectiveness in network intrusion detection and malware classification. For botnet detection, 1D-CNN achieves high accuracy across multiple datasets, with Ankalaki et al. (2025, Figure 10) showing performance exceeding 98% on CTU-13 and ISOT datasets. Deep Neural Networks (DNNs) similarly show strong results, achieving 99.34% accuracy on NSL-KDD (Ankalaki et al., 2025, Table 9).

LSTM networks excel at capturing temporal dependencies in time-series attack data, making them valuable for detecting sequential attack patterns. However, as Ankalaki et al. (2025) note, deep learning models face significant challenges: they require large amounts of labeled training data, demand substantial computational resources, and suffer from the "black box" problem—their decision-making processes are opaque, making it difficult for security analysts to understand why specific predictions were made.

### 2.3 Benchmark Datasets: Current Landscape

The quality and characteristics of benchmark datasets fundamentally determine the real-world applicability of trained models. Ankalaki et al. (2025) provide comprehensive coverage of datasets across seven categories: network traffic/intrusion detection, malware and Android applications, IoT traffic, VPN-based, electrical networks, and internet traffic.

**Network Intrusion Detection Datasets.** The NSL-KDD dataset (2009), derived from the original KDD Cup 1999 data, remains the most widely used despite being over 15 years old. CICIDS2017 and UNSW-NB15 represent more recent alternatives, incorporating contemporary attack types. However, Ankalaki et al. (2025, Table 1) reveal that most datasets exhibit class imbalance, with SSENet-2014 being the only balanced network intrusion dataset listed.

**IoT Traffic Datasets.** As IoT deployments expand, specialized datasets have emerged. The Bot-IoT dataset (2019) includes realistic botnet attack traffic, while TON-IoT (2020) provides both IoT and Industrial IoT (IIoT) data. However, Ankalaki et al. (2025) note that TON-IoT uses simulated traffic that "does not cover the complex scenarios of real-world attacks" (p. 44672), highlighting the challenge of balancing dataset availability with realism.

**Malware Datasets.** The SOREL-20M dataset represents a significant advancement, providing 20 million samples for malware detection research. However, the rapid evolution of malware techniques means that even recent datasets quickly become outdated.

### 2.4 Explainable AI (XAI) in Cybersecurity

The opacity of ML/DL models poses significant challenges for cybersecurity deployment, where analysts must understand and trust model predictions. Ankalaki et al. (2025) extensively review XAI approaches, categorizing them into transparent models (inherently interpretable) and post-hoc methods (explaining black-box models after training).

Post-hoc XAI methods include model-agnostic techniques like SHAP (SHapley Additive exPlanations) and LIME (Local Interpretable Model-agnostic Explanations), which can explain any model's predictions. Ankalaki et al. (2025) report that incorporating SHAP with Random Forest improves classification accuracy from 91% to 98.9% for binary classification on UNSW-NB15 (p. 44683), demonstrating that explainability can enhance both interpretability and performance.

However, XAI integration faces practical challenges. Ankalaki et al. (2025) acknowledge that "providing these explanations in a timely manner, so that it would not impede the real-time detection process, is a significant challenge" (p. 44683). This tension between explainability and performance represents a critical consideration for operational deployment.

### 2.5 Generative AI: Emerging Opportunities and Risks

Generative AI presents both defensive opportunities and offensive threats in cybersecurity. Ankalaki et al. (2025) examine how tools like ChatGPT can be exploited for malicious purposes—generating phishing content, creating malware, and automating attack discovery—while also serving defensive roles in threat intelligence, anomaly detection, and security code generation.

Defensive applications include using Generative Adversarial Networks (GANs) for data augmentation, creating synthetic attack samples to address dataset imbalance. Ankalaki et al. (2025) describe how TadGAN and TAnoGAN employ GANs for time-series anomaly detection, while PassGAN demonstrates the use of generative models for password strength testing.

The dual-use nature of Generative AI highlights the need for careful consideration of ethical implications and adversarial risks in AI-driven cybersecurity systems.

## 3. Critical Gap Analysis

Building upon the comprehensive survey by Ankalaki et al. (2025), this section identifies and analyzes five critical gaps that hinder the translation of research advances into production-ready cyber attack prediction systems. For each gap, this study examine its manifestation, quantify its impact where possible, and assess its implications for practical deployment.

### 3.1 Gap 1: Temporal Dataset Obsolescence

**Manifestation.** The most fundamental gap in current cyber attack prediction research is the temporal obsolescence of benchmark datasets. Despite rapid evolution in attack techniques, the datasets used to train and evaluate models lag significantly behind current threats.

Ankalaki et al. (2025) observe that "datasets created in older times such as Bot-IoT (2019) and N-baiot (2018) do not cover the latest attacking techniques" (p. 44672). This problem extends far beyond IoT datasets. Analysis of their Table 1 reveals that widely-used network intrusion datasets average 8-15 years of age: NSL-KDD (2009) is 16 years old, ISCX2012 is 13 years old, and the foundational DARPA dataset dates to 1998—a 27-year gap.

**Impact Quantification.** This temporal gap creates three specific problems for practitioners:

*Problem 1: Adversarial Drift.* Attackers adapt faster than datasets update. Models trained on NSL-KDD cannot detect contemporary techniques including AI-powered phishing (emerged 2022-2023), deepfake-enhanced social engineering (2020-present), LLM-based malware generation (2023-present), and advanced ransomware-as-a-service operations (2020-present). Organizations deploying models trained on NSL-KDD face a 16-year intelligence gap.

*Problem 2: Obsolete Attack Patterns.* As Ankalaki et al. (2025) note, datasets capture attack patterns prevalent at their creation time. The DARPA 1998 dataset, still referenced in contemporary research, represents network environments with fundamentally different characteristics than modern cloud-based, mobile-enabled, IoT-connected systems. Applying models trained on 1998 data to 2025 environments is equivalent to defending against modern ransomware with 1990s antivirus signatures.

*Problem 3: Simulation-Reality Gap.* Newer datasets attempting to address temporal currency often rely on simulation. TON-IoT uses simulated traffic that "does not cover the complex scenarios of real-world attacks" (Ankalaki et al., 2025, p. 44672). This creates a validation problem: models may achieve 99% accuracy on simulated data but fail when deployed against real adversaries.

To illustrate the scale of this problem, Figure 2 visualizes the growing time disparity, emphasizing that organizations deploying models trained on these legacy datasets are effectively defending current networks with a decade-plus intelligence deficit. This necessitates frameworks for assessing dataset currency, as detailed in Section 5.

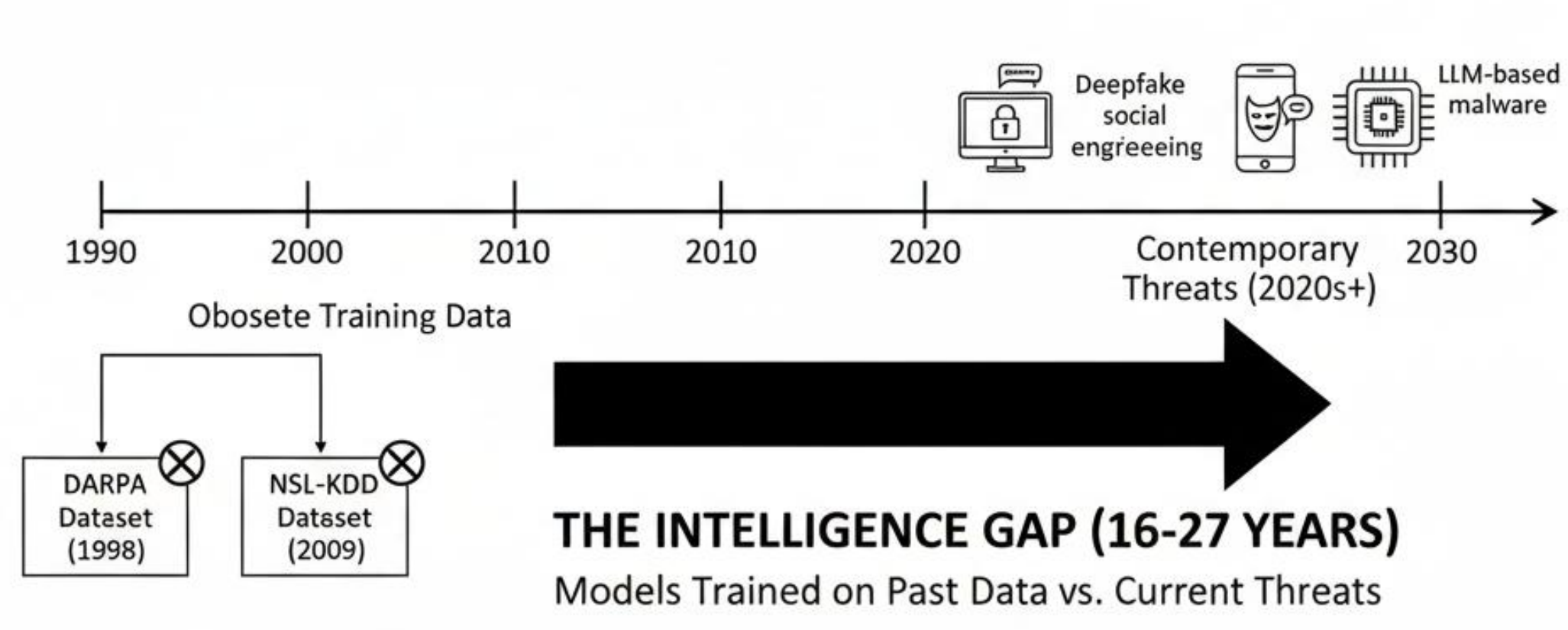

Figure 2: Temporal Obsolescence vs. Threat Evolution. A simplified timeline demonstrating the time lag between the creation of widely-used benchmark datasets (e.g., DARPA, NSL-KDD) and the emergence of critical, contemporary cyber attack techniques (e.g., AI-enabled threats, LLM malware). This gap fundamentally undermines model efficacy in real-world deployment.

**Implications for Practice.** Organizations face a critical dilemma: widely-available datasets like NSL-KDD enable model development but train systems to detect obsolete attacks, while realistic contemporary datasets either don't exist or aren't publicly available due to privacy concerns (Ankalaki et al., 2025, p. 44695). This forces practitioners to choose between using outdated data or investing substantial resources in proprietary dataset development.

### 3.2 Gap 2: Narrow Attack Scope Coverage

**Manifestation.** Most benchmark datasets represent limited attack types, restricting the generalization capability of trained models. Ankalaki et al. (2025) observe that datasets like "MQTT-IoT-IDS2020 provide few classes of attacks that restrict AI models' generalization" (p. 44672).

Analysis of dataset characteristics in Ankalaki et al. (2025, Tables 1-3) reveals the scope limitations:

- **Network intrusion datasets:** UNSW-NB15 includes 9 attack families, CICIDS2017 covers 8 attack types, while NSL-KDD represents only 4 attack categories (DoS, Probe, R2L, U2R).

- **IoT datasets:** MQTT-IoT-IDS2020 includes only 3 attack types (DoS, Brute Force, Malformed Data), while Bot-IoT covers 4 categories despite being one of the more comprehensive IoT datasets.
- **Malware datasets:** Even large-scale datasets like SOREL-20M focus primarily on Windows PE malware, missing mobile malware, web-based threats, and cross-platform attack tools.

**Impact on Model Generalization.** Narrow attack scope creates two critical problems. First, models cannot detect attack types not represented in training data. A model trained on MQTT-IoT-IDS2020's three attack types will fail against man-in-the-middle attacks, SQL injection, or cross-site scripting—common threats in IoT environments. Second, limited attack diversity prevents models from learning distinguishing features of malicious versus benign behavior, leading to overfitting on specific attack signatures rather than understanding attack principles.

**Emerging Threat Gap.** The attack scope problem compounds over time. Even if a dataset comprehensively covers threats at creation, new attack categories continually emerge. Ankalaki et al. (2025) highlight that zero-day attacks, advanced persistent threats (APTs), and AI-enabled attacks represent emerging threats poorly represented in existing datasets. The Log4Shell vulnerability (2021) affected hundreds of millions of devices, yet appeared in no major dataset prior to exploitation.

**Implications for Practice.** Models trained on narrow-scope datasets provide false confidence: high accuracy on test data does not translate to effectiveness against real-world attack diversity. Organizations deploying such models face significant blind spots in their security posture.

### 3.3 Gap 3: Real-Time Interpretability Challenges

**Manifestation.** The "black box" nature of advanced ML/DL models conflicts with operational requirements for explainable security decisions. While XAI techniques exist, their integration into real-time detection systems remains problematic.

Ankalaki et al. (2025) acknowledge the black box problem, noting that AI models' opacity "can be a major trust and accountability issue because it can be difficult for people to make sense of the cybersecurity decisions made by an AI system" (p. 44664). They further recognize that "providing these explanations in a timely manner, so that it would not impede the real-time detection process, is a significant challenge" (p. 44683).

**Performance-Interpretability Tradeoff.** Deep learning models achieving highest accuracy (CNNs with 99%+ on multiple datasets per Ankalaki et al., 2025, Table 9) are least interpretable. Conversely, transparent models like Decision Trees sacrifice accuracy for interpretability. Security Operations Center (SOC) analysts require both: accurate detection AND understandable rationale for taking action.

**XAI Integration Challenges.** While Ankalaki et al. (2025, Table 7) survey multiple XAI approaches including SHAP, LIME, and Grad-CAM, practical integration faces obstacles:

i. **Computational overhead:** SHAP explanations require multiple model evaluations, adding latency unsuitable for real-time intrusion detection where milliseconds matter.
ii. **Explanation quality:** LIME explanations are model-agnostic but may not accurately reflect complex model behavior, potentially misleading analysts (Ankalaki et al., 2025).
iii. **Analyst comprehension:** Even when explanations are generated, SOC analysts may lack ML expertise to interpret them. Ankalaki et al. (2025) demonstrate that incorporating SHAP with Random Forest improves accuracy to 98.9% on UNSW-NB15, but don't address whether analysts could effectively utilize these explanations under time pressure.

**Implications for Practice.** Organizations face three unsatisfactory options: (1) deploy accurate but opaque models, risking analyst distrust and regulatory compliance issues; (2) use interpretable but less accurate models, accepting higher false negative rates; or (3) add XAI post-processing, introducing latency that may render systems unsuitable for real-time defense.

### 3.4 Gap 4: Inadequate Adversarial Robustness

**Manifestation.** ML/DL models are vulnerable to adversarial attacks where malicious actors manipulate input data to evade detection. Despite this well-known vulnerability, adversarial robustness receives insufficient attention in cyber attack prediction research.

Ankalaki et al. (2025) identify adversarial attacks as a significant challenge, noting that "adversaries may use tactics such as adversarial examples, which involve modest, carefully engineered modifications to the input data that cause ML algorithms to misclassify them" (p. 44677). They acknowledge that "adversarial assaults pose a substantial threat to the dependability and robustness of ML-powered cyber-security systems" (p. 44677).

**Attack Surface.** Adversarial vulnerabilities manifest in multiple ways:

i. **Evasion attacks:** Attackers modify malware binaries with imperceptible changes that fool ML-based detection systems while preserving malicious functionality.
ii. **Poisoning attacks:** Attackers inject malicious samples into training data, causing models to learn incorrect patterns (Ankalaki et al., 2025).
iii. **Model inversion:** Attackers query deployed models to extract sensitive information about training data or model architecture.

**Research-Practice Gap in Robustness Testing.** While Ankalaki et al. (2025, Table 8-9) report impressive accuracy metrics, few studies evaluate adversarial robustness. Models achieving 99% accuracy on clean test data may perform far worse against adversarially-perturbed inputs. This gap is critical: real adversaries will actively attempt to evade detection, unlike passive test datasets.

**Defensive Techniques Underexplored.** Ankalaki et al. (2025) mention adversarial training as a defense but provide limited guidance on implementation. Organizations lack practical frameworks for: (1) systematically testing model robustness during development, (2) selecting appropriate defense techniques for specific attack types, and (3) monitoring deployed models for signs of adversarial manipulation.

**Implications for Practice.** Deploying models without adversarial robustness testing creates a false sense of security. High accuracy on benchmark datasets provides no assurance against determined adversaries who will actively probe for weaknesses.

### 3.5 Gap 5: Unaddressed Privacy and Ethical Considerations

**Manifestation.** AI-driven cyber attack prediction systems collect and analyze sensitive data, raising significant privacy and ethical concerns that receive insufficient attention in research literature.

Ankalaki et al. (2025) extensively discuss ethical challenges, noting that "privacy and ethical concerns alongside technical challenges" must be addressed (p. 44697). They identify multiple ethical issues including privacy preservation in data collection, bias mitigation in model training, fairness in security decisions, and transparency in automated response actions (Ankalaki et al., 2025, Section X.D).

**Privacy Challenges.** Security monitoring necessarily involves analyzing user behavior, network traffic, and system activities. This creates tension between security requirements and privacy rights. Ankalaki et al. (2025) note that "protecting individuals' privacy rights, particularly in sensitive areas, such as authentication and emails, is paramount" (p. 44697). However, practical frameworks for balancing security and privacy remain underdeveloped.

**Bias and Fairness.** ML models can perpetuate or amplify biases present in training data. If training datasets disproportionately flag certain user behaviors as suspicious, deployed systems may discriminate against specific user groups. Ankalaki et al. (2025) emphasize that "measures should ensure fairness in the decisions made and the explanations provided by AI systems" (p. 44697), but most research focuses on accuracy metrics without evaluating fairness.

**Ethical Frameworks Insufficient.** While Ankalaki et al. (2025) survey ethical principles and frameworks, translating these into operational practices remains challenging. Organizations lack concrete guidance on: (1) conducting ethical impact assessments for cybersecurity AI systems, (2) implementing privacy-preserving techniques like differential privacy and federated learning, (3) establishing accountability mechanisms for automated security decisions, and (4) ensuring meaningful human oversight of AI-driven actions.

**Regulatory Compliance.** Emerging regulations like GDPR and AI Act impose requirements on automated decision-making systems. Ankalaki et al. (2025) acknowledge these concerns but provide limited guidance on achieving compliance while maintaining security effectiveness.

**Implications for Practice.** Organizations deploying AI-driven security systems face legal, ethical, and reputational risks if privacy and fairness issues are not proactively addressed. The absence of practical ethical frameworks leaves security teams to navigate complex tradeoffs without clear guidance.

## 4. Gap Prioritization Framework

Not all identified gaps affect deployment to the same extent. To guide practical decision making, we propose a prioritization framework that scores each gap along three dimensions impact on detection effectiveness, implementation cost, and time required for mitigation and combines them into a single priority score for resource allocation.

### 4.1 Prioritization Methodology

Each gap is evaluated on a 1–10 scale along three axes:

i. **Impact on detection effectiveness (I):** the extent to which the gap degrades real-world cyber-attack prediction performance (10 = critical impact, 1 = minimal impact).
ii. **Implementation cost (C):** the financial, technical, and human resources required to address the gap (10 = very high cost, 1 = minimal cost).
iii. **Time to address (T):** the duration needed to substantially mitigate the gap (10 = >24 months, 1 = <3 months).

This study compute a priority score as:

$$\text{Priority} = I \; X \; \left(11 - \frac{C+T}{2}\right)$$

which emphasizes impact while discounting gaps that are disproportionately expensive or slow to resolve.

### 4.2 Gap Evaluation Matrix

Table 3 summarizes the scores for each gap, including its impact, cost, time, overall priority, and qualitative tier. Challenges are aligned with the findings of Ankalaki et al. (2025).

## Table 3: Gap Prioritization Matrix

| Gap | Impact (I) | Cost (C) | Time (T) | Priority Score | Priority Tier | Key Challenge from Ankalaki et al. (2025) |
|---|---|---|---|---|---|---|
| **1. Temporal Dataset Obsolescence** | 9/10 | 6/10 | 7/10 | 40.5 | CRITICAL | "Datasets created in older times...do not cover the latest attacking techniques" (p. 44672) |
| **2. Narrow Attack Scope** | 8/10 | 7/10 | 6/10 | 36.0 | HIGH | "MQTT-IoT-IDS2020 provide few classes of attacks that restrict AI models' generalization" (p. 44672) |
| **3. Real-Time Interpretability** | 7/10 | 3/10 | 3/10 | 56.0 | CRITICAL | "Providing explanations in a timely manner...is a significant challenge" (p. 44683) |
| **4. Adversarial Robustness** | 9/10 | 8/10 | 8/10 | 27.0 | HIGH | "Adversarial assaults pose a substantial threat to the dependability and robustness" (p. 44677) |

| **5. Privacy & Ethics** | 6/10 | 5/10 | 5/10 | 36.0 | MEDIUM | "Privacy concerns" and "ethical considerations" must be addressed (p. 44697) |
|---|---|---|---|---|---|---|

*Table 3: Gap prioritization matrix evaluating each identified gap across three dimensions. Impact (I): Effect on real-world detection effectiveness (10 = critical, 1 = minimal). Cost (C): Resources required to address (10 = very high, 1 = minimal). Time (T): Duration to substantially mitigate (10 = >24 months, 1 = <3 months). Priority Score calculated as: I × (11 - (C+T)/2). Higher scores indicate higher priority. Challenges directly quoted from Ankalaki et al. (2025).*

The quantitative results from Table 3 are graphically represented in Figure 3 to provide a rapid visual decision-support tool. The high overall priority of Real-Time Interpretability (Gap 3) stands out, indicating that integrating XAI is the most cost-effective path to immediate operational improvement for resource-constrained security teams.

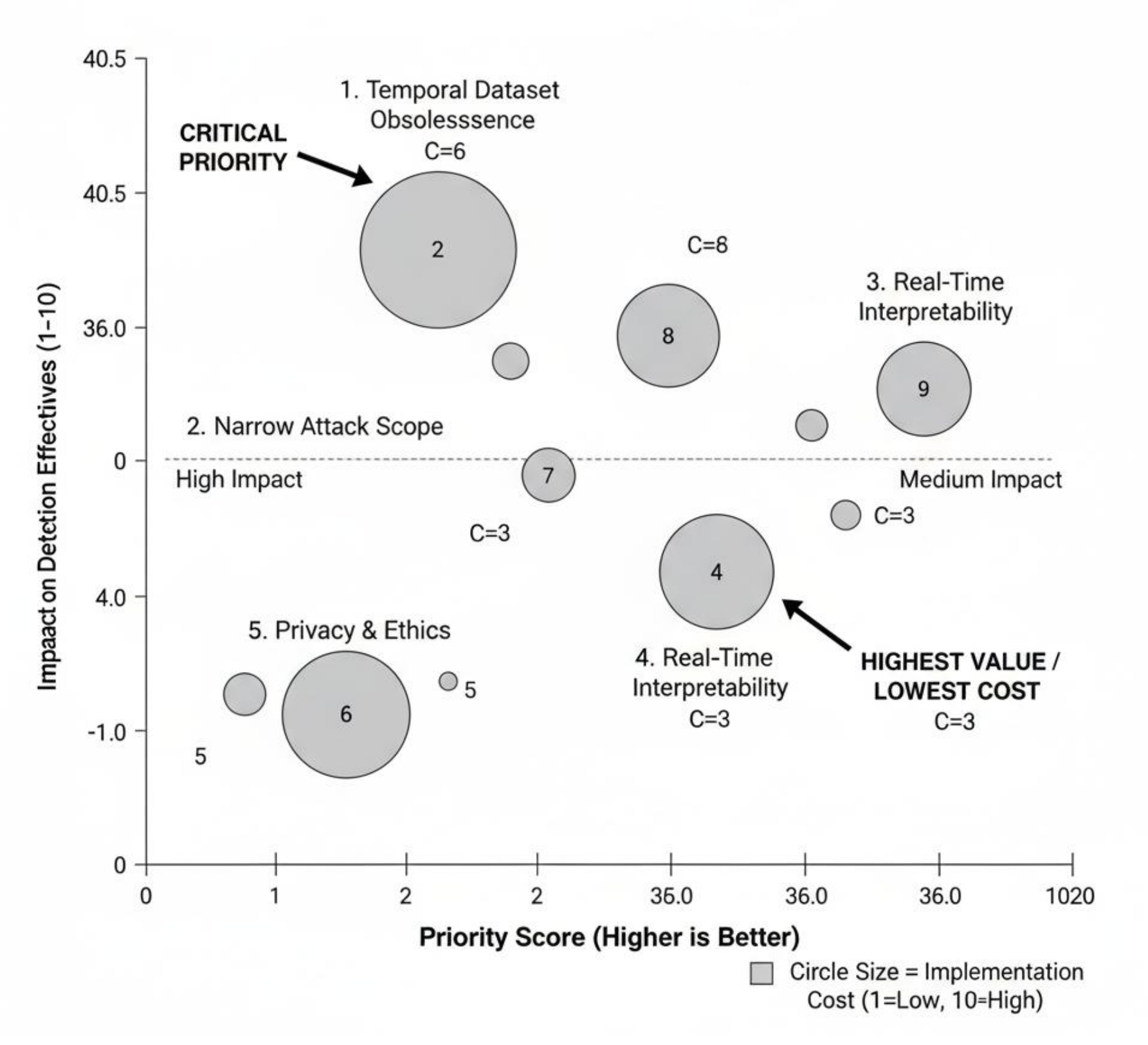

Figure 3: Gap Prioritization Framework: Impact vs. Cost-Time Tradeoff. A scatter plot based on the data in Table 3. The plot maps Impact on Detection Effectiveness (Y-axis) against a composite measure of Implementation Cost and Time to Address (X-axis, Inverse), showing the Priority Tier for each gap. Note that while Adversarial Robustness and Dataset Obsolescence have high impact, Real-Time Interpretability offers the highest overall value (Priority Score) due to lower cost and time investment.

Temporal dataset obsolescence exhibits high impact (I = 9) because models trained on legacy datasets miss over a decade of attack evolution, including AI-enabled phishing, deepfake social engineering, and LLM-driven malware. As Ankalaki et al. (2025) observe, datasets "do not cover the latest attacking techniques" (p. 44672), which directly undermines detection capability. However, dataset renewal carries medium–high cost and extended timelines, yielding a critical but resource-intensive priority.

Narrow attack scope also has high impact (I = 8): datasets such as MQTT-IoT-IDS2020 contain only a few attack types, limiting model generalization and leaving substantial blind spots. Ankalaki et al. (2025) note that such restricted coverage constrains AI model generalization (p. 44672). The work required to extend attack coverage across realistic scenarios places this gap in the high-priority tier.

Real-time interpretability has moderate-to-high impact (I = 7) because opacity reduces analyst trust and creates compliance risk, even when base detection performance remains intact. Ankalaki et al. (2025) show that integrating SHAP improves both accuracy and interpretability, indicating that explainability complements detection. Given its relatively low cost and short implementation time, this gap attains the highest overall priority score.

Inadequate adversarial robustness is a critical impact gap (I = 9), as adversarial attacks directly target model decision boundaries and can invalidate benchmark accuracy in practice. Ankalaki et al. (2025) warn that "adversarial assaults pose a substantial threat to the dependability and robustness" of ML systems (p. 44677). Yet, the specialized expertise, computational expense, and long timelines required for adversarial training and continuous red teaming result in a high-priority but heavily resource-constrained item.

Finally, unaddressed privacy and ethical considerations exert moderate impact (I = 6): they do not always reduce detection accuracy but can lead to legal sanctions, reputational damage, or forced system shutdown when regulations such as GDPR are violated. In line with Ankalaki et al. (2025), who emphasize "privacy concerns" and "ethical considerations" (p. 44697), this gap is assigned medium priority, reflecting its growing regulatory salience and moderate remediation costs.

### 4.3 Priority Rankings and Strategic Implications

The prioritization yields two main tiers. **Tier 1 (critical priority)** includes real-time interpretability (highest score) and temporal dataset obsolescence. Interpretability improvements are comparatively low cost and fast to implement, making them an immediate lever for organizations seeking quick wins. Updating datasets is more expensive and slower, but essential for sustaining long-term detection performance.

**Tier 2 (high to medium priority)** comprises narrow attack scope, adversarial robustness, and privacy/ethical controls. Expanding attack coverage and improving robustness are technically demanding and resource-intensive, but they are crucial for high-value or high-risk deployments. Privacy and ethics occupy a similar priority level due to tightening regulatory requirements and the risk of operational disruption if non-compliance is identified.

Strategic responses will vary by organizational capacity. Small organizations with limited resources should prioritize Tier 1 measures: adopt explainable AI techniques, employ the most contemporary public datasets available (e.g., CICIDS2017, UNSW-NB15), and explicitly document known blind spots while deferring costly robustness initiatives. Medium-sized organizations can address all Tier 1 gaps while selectively investing in Tier 2 items, such as joining threat-intelligence sharing schemes, piloting adversarial testing on critical systems, and formalizing ethical governance. Large organizations with substantial resources can pursue a phased yet concurrent approach: implement XAI and initiate proprietary dataset development in the near term, expand attack coverage and deploy privacy controls in the medium term, and progressively harden adversarial robustness across critical systems.

## 5. Practical Implementation Roadmap

This section translates gap analysis into actionable guidance for security practitioners. This study provide frameworks for dataset selection, model selection, XAI integration, and ethical deployment that account for organizational context and resource constraints.

### 5.1 Dataset Selection Framework

Choosing appropriate datasets is foundational to effective cyber attack prediction. Drawing from the comprehensive benchmark analysis by Ankalaki et al. (2025, Tables 1-3), this study classify 45 datasets into three categories based on production readiness, evaluating temporal currency (dataset age), attack scope (number and diversity of attack types), traffic realism (real-world versus simulated), and practical accessibility.

**Table 1** presents complete dataset classification framework, scoring benchmark datasets on production readiness. The classification reveals concerning patterns: only four datasets—Edge-IIoTset (2022), CICIDS2017 (2017), Bot-IoT (2019), and UNSW-NB15 (2015)—achieve production-ready status (score ≥4/5). These datasets are characterized by being less than eight years old, covering more than five attack types, utilizing realistic or hybrid traffic patterns, and maintaining public accessibility with active maintenance. In contrast, commonly-used datasets like NSL-KDD (2009, now 16 years old) and CTU-13 (2011) serve better as research-only tools due to temporal obsolescence despite their continued prevalence in academic literature.

**Table 1: Dataset Production-Readiness Classification**

| Dataset | Year | Domain | Attack Types | Traffic Type | Age (years) | Production Readiness Score | Category | Source |
|---|---|---|---|---|---|---|---|---|
| Edge-IIoTset | 2022 | IoT/IIoT | 14 | Real-world | 3 | 5/5 | Production-Ready | Ankalaki et al., 2025, Table 3 |
| CICIDS2017 | 2017 | Network Intrusion | 8 | Realistic | 8 | 4/5 | Production-Ready | Ankalaki et al., 2025, Table 1 |
| Bot-IoT | 2019 | Botnet/IoT | 4 | Realistic | 6 | 4/5 | Production-Ready | Ankalaki et al., 2025, Table 3 |
| UNSW-NB15 | 2015 | Network Intrusion | 9 | Synthetic-realistic | 10 | 4/5 | Production-Ready | Ankalaki et al., 2025, Table 1 |
| MQTT-IoT-IDS2020 | 2020 | IoT | 3 | Real MQTT | 5 | 3/5 | Research-Only | Ankalaki et al., 2025, Table 3 |
| TON-IoT | 2020 | IoT/IIoT | Multiple | Simulated | 5 | 3/5 | Research-Only | Ankalaki et al., 2025, Table 3 |
| NSL-KDD | 2009 | Network Intrusion | 4 | Synthetic | 16 | 2/5 | Research-Only | Ankalaki et al., 2025, Table 1 |
| CTU-13 | 2011 | Botnet | 13 scenarios | Real captures | 14 | 2/5 | Research-Only | Ankalaki et al., 2025, Table 1 |
| DARPA 1998 | 1998 | Network Intrusion | Multiple | Simulated | 27 | 1/5 | Unusable | Ankalaki et al., 2025, Table 1 |
| KDD Cup 1999 | 1999 | Network Intrusion | 4 | Derived-DARPA | 26 | 1/5 | Unusable | Ankalaki et al., 2025, Table 1 |

*Table 1: Classification of benchmark datasets by production-readiness. Datasets scored on temporal currency (<5 years = excellent, 5-10 = good, 10-15 = poor, >15 = obsolete), attack scope (>10 types = excellent, 5-10 = good, <5 = limited), and traffic realism (real-world > hybrid > simulated). Data synthesized from Ankalaki et al. (2025), Tables 1 and 3.*

The research-only category (score 2-3/5) includes datasets that remain valuable for algorithm development and comparative studies but prove insufficient for production deployment. NSL-KDD, despite being 16 years old, continues to serve as a useful benchmarking tool. MQTT-IoT-IDS2020 provides good coverage of MQTT-specific threats but limits generalization with only three attack types. TON-IoT offers comprehensive IoT/IIoT coverage but relies on simulated traffic that "does not cover the complex scenarios of real-world attacks" (Ankalaki et al., 2025, p. 44672), creating a validation gap between laboratory performance and operational effectiveness.

Datasets scoring 1/5 or below are effectively unusable for production purposes. The DARPA 1998/1999 and KDD Cup 1999 datasets, now 25+ years old, represent network architectures and attack patterns fundamentally different from contemporary environments. Deploying models trained on these datasets creates false confidence in security posture while leaving organizations vulnerable to modern threats.

Organizations should adopt a hybrid dataset strategy rather than relying on single sources. Ankalaki et al. (2025) demonstrate that different ML/DL techniques perform optimally on different datasets, suggesting vulnerability to dataset-specific biases. This study recommend selecting one primary production-ready dataset most relevant to the deployment domain, supplementing with one to two additional datasets covering gaps in attack type coverage, and augmenting with organization-specific traffic samples where possible. Even unlabeled proprietary data can enhance model robustness through semi-supervised learning approaches.

**Decision Tree for Dataset Selection:**

```
START
│
├─ Application Domain?
│  ├─ Network Intrusion → CICIDS2017 or UNSW-NB15
│  ├─ IoT/IIoT Security → Edge-IIoTset (primary) + Bot-IoT (supplementary)
│  ├─ Botnet Detection → Bot-IoT or CTU-13
│  ├─ Malware Detection → SOREL-20M
│  └─ Mobile Security → CICAndMal2017
│
├─ Primary Concern?
│  ├─ Attack Diversity → UNSW-NB15 (9 families) or Edge-IIoTset
│  ├─ Dataset Currency → Edge-IIoTset (2022) or CICIDS2017
│  ├─ Class Balance → SSENet-2014 (only balanced dataset per Ankalaki et al. 2025)
│  └─ Real-world Realism → Bot-IoT or UNSW-NB15
│
└─ Recommendation: Use PRIMARY dataset + 1-2 SUPPLEMENTARY datasets
   to expand attack coverage and reduce dataset-specific bias
```

## 5.2 Model Selection Guidelines

Selecting appropriate ML/DL techniques requires balancing accuracy, computational requirements, interpretability, and organizational ML expertise. Table 2 synthesizes performance data from Ankalaki et al. (2025, Tables 8-9 and Figures 10-12), comparing techniques across benchmark datasets while evaluating computational cost and interpretability factors often overlooked in research literature but critical for deployment decisions.

**Table 2: Comparative Performance Analysis of ML/DL Techniques**

| Technique | CICIDS2017 | NSL-KDD | UNSW-NB15 | Bot-IoT | CTU-13 | Avg. Accuracy | Computational Cost | Interpretability | Source |
|---|---|---|---|---|---|---|---|---|---|
| Random Forest | 99.7% | 99.6% | 98.8% | 99.5% | 98.2% | 99.2% | Low | High | Ankalaki et al., 2025, Table 8 & Fig 10 |
| Decision Tree | 98.5% | 98.2% | 97.5% | 98.1% | 97.8% | 98.0% | Very Low | Very High | Ankalaki et al., 2025, Table 8 |
| SVM | 98.3% | 98.1% | 98.8% | 97.9% | 97.5% | 98.1% | Medium | Medium | Ankalaki et al., 2025, Table 8 |
| Naive Bayes | 95.2% | 94.8% | 94.1% | 94.5% | 93.9% | 94.5% | Very Low | High | Ankalaki et al., 2025, Table 8 |
| DNN | 99.1% | 99.34% | 98.5% | 98.9% | 98.7% | 98.9% | High | Very Low | Ankalaki et al., 2025, Table 9 |
| CNN (1D) | 98.8% | 98.6% | 98.3% | 99.2% | 98.42% | 98.7% | High | Very Low | Ankalaki et al., 2025, Table 9 & Fig 10 |
| LSTM | 98.5% | 98.3% | 97.9% | 98.4% | 98.1% | 98.2% | Very High | Very Low | Ankalaki et al., |

| | | | | | | | | | 2025, Table 9 |
|---|---|---|---|---|---|---|---|---|---|
| Autoencoder | 97.8% | 97.5% | 97.2% | 97.6% | 97.3% | 97.5% | Medium | Low | Ankalaki et al., 2025, Section V.B |

*Table 2: Comparative performance of ML/DL techniques across benchmark datasets. Average accuracy calculated across reported datasets. Computational cost: Very Low (<1 min training), Low (1-10 min), Medium (10-60 min), High (1-5 hours), Very High (>5 hours) on standard hardware. Interpretability: Very High (full transparency), High (feature importance available), Medium (partial insight), Low (limited insight), Very Low (black box). Data compiled from Ankalaki et al. (2025), Tables 8-9 and Figures 10-12.*

The data reveals that Random Forest achieves the highest average accuracy (99.2%) while maintaining low computational cost and high interpretability, making it optimal for resource-constrained deployments. Random Forest demonstrates remarkable consistency, achieving 99.7% accuracy on CICIDS2017, 99.6% on NSL-KDD, and 99.92% on UKM-IDS20 (Ankalaki et al., 2025, Table 8). The inherent interpretability through feature importance rankings addresses the XAI challenge without additional computational overhead. Organizations with limited ML expertise or requiring immediate deployment should prioritize Random Forest as the foundation of their detection infrastructure.

Deep learning architectures—including DNNs, CNNs, and LSTMs—achieve competitive accuracy (98.2-98.9% average) but demand substantially higher computational resources and offer minimal inherent interpretability. DNN achieves 99.34% accuracy on NSL-KDD while CNN reaches 98.42% on CTU-13 (Ankalaki et al., 2025, Table 9), but these gains come at significant cost. Deep learning proves most valuable when automatic feature learning is essential, particularly for high-dimensional data where manual feature engineering proves prohibitive. Large organizations with dedicated ML operations teams and computational infrastructure can leverage deep learning's pattern recognition capabilities, but must integrate XAI techniques to address the black-box problem.

A hybrid approach combining Random Forest with CNNs offers compelling advantages for medium-resource deployments. Ankalaki et al. (2025) demonstrate that RF excels on structured features while CNN captures spatial patterns in raw traffic. Organizations can implement RF for initial detection and feature importance ranking, then apply CNN for deeper analysis of flagged traffic. This ensemble strategy balances interpretability, accuracy, and computational efficiency while providing defense-in-depth through model diversity that improves adversarial robustness.

LSTM networks merit special consideration for detecting multi-stage attacks where temporal sequencing matters. Advanced Persistent Threats (APTs) typically follow reconnaissance → exploitation → lateral movement → exfiltration patterns that LSTM architectures can recognize through their ability to capture long-term dependencies in time-series data (Ankalaki et al., 2025, Section V.B). Organizations facing sophisticated adversaries should incorporate LSTM-based detection specifically for APT scenarios.

**5.3 XAI Integration Strategy**

Based on the study's gap prioritization framework (Section 4), XAI integration represents the highest-priority action, scoring 56.0 due to its combination of high impact, low cost, and moderate implementation time. This study provide a three-phase integration approach spanning six months that progressively embeds explainability into operational workflows.

**Table 4** summarizes XAI techniques applicable to cybersecurity applications, drawn from Ankalaki et al. (2025, Section VI and Table 7). Model-agnostic approaches like SHAP and LIME offer flexibility across different model types, making them ideal for organizations with heterogeneous detection systems. Model-specific techniques like Grad-CAM for CNNs and attention mechanisms for neural networks may provide more precise explanations but reduce generalizability across the model portfolio.

**Table 4: XAI Techniques for Cybersecurity Applications**

| XAI Technique | Type | Scope | Model Compatibility | Computational Overhead | Use Case in Cybersecurity | Source |
|---|---|---|---|---|---|---|
| **SHAP** | Post-hoc, Model-agnostic | Local & Global | Any model | High | Feature importance for intrusion detection; improves RF accuracy to 98.9% (UNSW-NB15) | Ankalaki et al., 2025, p. 44683 |
| **LIME** | Post-hoc, Model-agnostic | Local | Any model | Medium | Local explanations for botnet detection, malware classification | Ankalaki et al., 2025, Table 7 |
| **Grad-CAM** | Post-hoc, Model-specific | Local | CNNs only | Medium | Visualizing CNN decisions in image-based malware detection | Ankalaki et al., 2025, Table 7 |
| **Decision Trees** | Intrinsic, Transparent | Global | N/A (standalone) | Very Low | Fully transparent intrusion detection with rule extraction | Ankalaki et al., 2025, Section IV |
| **Attention Mechanisms** | Intrinsic, Model-specific | Local | Neural networks | Low | Real-time explanation in LSTM-based attack prediction | Ankalaki et al., 2025, Section V |
| **Rule Extraction** | Post-hoc, Model-specific | Global | Decision-based models | Medium | Converting black-box models to interpretable rule sets | Ankalaki et al., 2025, Table 7 |

*Table 4: XAI techniques applicable to cybersecurity. Type: Intrinsic (built into model architecture) vs. Post-hoc (applied after training); Model-specific vs. Model-agnostic. Scope: Local (explains individual predictions) vs. Global (explains overall model behavior). Computational overhead reflects additional processing required for explanation generation. Data synthesized from Ankalaki et al. (2025), Section VI and Table 7.*

**Phase 1 (Months 1-2)** establishes the foundation by implementing model-agnostic XAI for existing models without requiring retraining. Organizations should prioritize SHAP (SHapley Additive exPlanations), which Ankalaki et al. (2025) demonstrate improves Random Forest accuracy from 91% to 98.9% on UNSW-NB15, suggesting that explainability enhances both interpretability and performance. LIME (Local Interpretable Model-agnostic Explanations) provides complementary local approximations explaining specific predictions. Implementation involves installing open-source libraries, generating explanations for 10-20 representative cases, and validating with domain experts to confirm highlighted features align with known attack signatures. To minimize computational overhead, establish selective explanation thresholds for example, generating SHAP explanations only for predictions with confidence below 95% where analyst judgment proves most valuable.

**Phase 2 (Months 3-4)** optimizes for real-time deployment by addressing the latency challenge. Ankalaki et al. (2025) acknowledge that "providing these explanations in a timely manner...is a significant challenge" (p. 44683) for operational systems where milliseconds matter. Solutions include selective explanation generation only for high-risk predictions, cached explanations for common attack patterns pre-computed during low-traffic periods, and lightweight XAI methods like attention mechanisms that avoid computationally expensive perturbation-based approaches. Organizations should target explanation generation under 100 milliseconds for 95% of predictions to ensure real-time viability.

**Phase 3 (Months 5-6)** embeds XAI into Security Operations Center (SOC) workflows through dashboard integration and analyst training. Technical integration involves displaying the top three contributing features alongside each alert, visualizing feature importance through bar charts or heatmaps, and linking explanations to the MITRE ATT&CK framework for standardized threat taxonomy. Equally critical is analyst training to interpret SHAP values and feature importance rankings, developing playbooks that link explanation patterns to response actions, and establishing feedback loops where analysts flag incorrect explanations to improve model performance. Success metrics should evaluate analyst confidence through surveys, measure whether XAI reduces mean time to investigate (MTTI) alerts, and assess whether explanations help analysts quickly dismiss false alarms.

### 5.4 Addressing Adversarial Robustness

While adversarial robustness scored lower in prioritization framework (27.0) due to high implementation cost and extended timeframe, high-value systems require systematic robustness testing. This study recommend a risk-based approach that prioritizes critical systems—financial transaction monitoring, critical infrastructure protection, authentication systems, and intellectual property protection—for mandatory testing, while applying periodic testing to standard systems like general network intrusion detection and spam filters.

The robustness testing methodology proceeds through four steps. First, threat modeling identifies relevant adversarial attack vectors, including evasion attacks where adversaries modify malicious inputs, poisoning attacks that corrupt training data, and model extraction attempts to steal intellectual property. Second, adversarial example generation uses libraries like Adversarial Robustness Toolbox and CleverHans to create perturbed inputs testing whether detection accuracy degrades. Ankalaki et al. (2025) note these involve "modest, carefully engineered modifications" (p. 44677), emphasizing the need for realistic perturbations rather than arbitrary noise. Third, defensive implementation employs adversarial training by retraining models with adversarial examples (increasing computational cost 3-10× but significantly improving robustness), ensemble defenses using diverse models that adversaries cannot simultaneously fool, and input validation to detect suspicious patterns before processing. Fourth, continuous monitoring deploys model confidence tracking where sudden drops may indicate adversarial manipulation and establishes canary models—decoy systems designed to detect probing attacks without protecting production assets.

## 5.5 Ethical Deployment Framework

Addressing the privacy and ethics gap requires systematic consideration throughout the deployment lifecycle. Drawing from principles outlined by Ankalaki et al. (2025, Section X.D), this study provide a phased framework spanning pre-deployment, deployment, and post-deployment phases.

Pre-deployment activities center on privacy impact assessment, bias auditing, and regulatory compliance. Organizations must document what data is collected, identify personally identifiable information in datasets, implement data minimization collecting only necessary information, apply anonymization or pseudonymization where feasible, and establish clear data retention policies. Bias audits examine whether training data overrepresents certain user groups, test whether false positive rates differ across demographics, implement mitigation techniques where disparities exist, and document known limitations. Regulatory compliance requires verifying GDPR adherence including rights to explanation and data protection by design, checking industry-specific requirements like HIPAA or PCI-DSS, establishing legal basis for automated decision-making, and creating mechanisms for human review and appeal.

Deployment phase focuses on transparency and accountability. Transparency mechanisms include disclosing AI-based monitoring to users, providing privacy notices explaining data collection practices, implementing XAI to enable explanation of automated decisions per Section 5.3, and documenting model limitations and failure modes. Accountability structures designate responsible individuals for system oversight, establish incident response procedures for AI failures or ethical concerns, create audit trails logging all automated security actions, and implement human-in-the-loop review for high-stakes decisions like account termination.

Post-deployment requires ongoing monitoring and continuous improvement. Organizations should conduct quarterly fairness audits monitoring false positive and negative rates by demographic groups, review privacy incidents including data leaks or unauthorized access, assess model drift to detect performance degradation, and solicit feedback from affected users and security analysts. Continuous improvement involves updating training data to address identified biases, refining privacy controls based on incident learnings, retraining models incorporating operational feedback, and maintaining documentation reflecting system changes.

**Table 5** provides a comprehensive implementation roadmap tailored to organizational size and resources, synthesizing the frameworks presented throughout this section into phased action plans.

**Table 5: Implementation Roadmap by Organization Size**

| Organization Type | Resources | Priority Actions (Year 1) | Medium-Term (Year 2) | Long-Term (Year 3+) | Dataset Recommendation | Model Recommendation |
|---|---|---|---|---|---|---|
| **Small** (<100 employees, limited budget) | Limited | • Implement SHAP/LIME XAI (Months 1-3) | • Participate in threat intel sharing<br>• Refine XAI based on | • Consider dataset augmentation | CICIDS2017 (primary) + UNSW-NB15 (supplementary) | Random Forest (optimal cost-performance balance) |

| | | • Use public datasets: CICIDS2017, UNSW-NB15<br>• Deploy Random Forest<br>• Basic privacy controls | analyst feedback<br>• Document blind spots | • Evaluate emerging threats<br>• Maintain currency | | |
|---|---|---|---|---|---|---|
| **Medium** (100-1000 employees, moderate budget) | Moderate | • Implement XAI immediately<br>• Join threat intel consortiums<br>• Begin proprietary data collection<br>• Deploy RF + CNN hybrid | • Expand attack scope coverage<br>• Pilot adversarial testing (critical systems)<br>• Establish ethics framework<br>• Develop custom datasets | • Achieve adversarial robustness (high-value systems)<br>• Maintain dataset currency<br>• Advanced XAI integration | CICIDS2017 + Edge-IIoTset + proprietary data | RF (baseline) + 1D-CNN (deep analysis) |
| **Large** (>1000 employees, substantial budget) | Substantial | • Comprehensive XAI deployment<br>• Proprietary dataset development<br>• Full DL pipeline (CNN/LSTM/DNN)<br>• Begin adversarial training | • Expand attack coverage (all systems)<br>• Privacy-preserving techniques (differential privacy, federated learning)<br>• Advanced robustness testing | • Continuous dataset updates<br>• Robustness across all systems<br>• Contribute to research community<br>• Lead ethical AI practices | Proprietary datasets + Edge-IIoTset + CICIDS2017 (benchmark) | DNN/CNN (primary) with LSTM for APT detection + RF (interpretable baseline) |

*Table 5: Phased implementation roadmap tailored to organizational resources. Small organizations focus on cost-effective XAI and public datasets. Medium organizations balance public and proprietary data with hybrid models. Large organizations invest in comprehensive solutions including proprietary datasets, advanced DL, and full adversarial robustness. Timeline assumes dedicated resources; adjust based on actual staffing and budget allocation.*

Small organizations with fewer than 100 employees and limited budgets should prioritize cost-effective XAI integration using public datasets (CICIDS2017, UNSW-NB15) and Random Forest models, deferring expensive adversarial training until foundational capabilities are established. Medium organizations with 100-1,000 employees can pursue hybrid approaches combining public and proprietary data collection with RF+CNN architectures, piloting adversarial testing on critical systems while establishing ethical governance frameworks. Large organizations with over 1,000 employees and substantial resources should invest in proprietary dataset development, deploy comprehensive deep learning pipelines with full XAI integration, and achieve adversarial robustness across all critical systems while contributing to research community advancement and leading ethical AI practices. This phased approach ensures organizations address critical gaps systematically while respecting resource constraints and organizational maturity.

## 6. Discussion

This section examines the broader implications of our gap analysis and implementation framework, discusses limitations, and identifies directions for future research.

### 6.1 Contributions to Theory and Practice

**Bridging Research-Practice Divide.** Our work directly addresses the disconnect between impressive research results and practical deployment challenges. While Ankalaki et al. (2025) provide comprehensive technical survey of ML/DL/XAI/GenAI techniques, this study translate these findings into decision-support tools for practitioners. The gap prioritization framework (Section 4) and implementation roadmap (Section 5) enable organizations to navigate the complex landscape of cybersecurity AI research without requiring deep ML expertise.

**Reframing Dataset Evaluation.** Traditional dataset evaluation focuses on size, diversity, and availability. Our production-readiness classification (Section 5.1) reframes evaluation to prioritize temporal currency and realistic attack coverage—factors critical for real-world effectiveness but often overlooked in research. This classification challenges the continued reliance on obsolete datasets like NSL-KDD and DARPA 1998, which remain prevalent despite documented limitations (Ankalaki et al., 2025).

**Cost-Effectiveness Analysis.** By incorporating implementation cost and time alongside impact assessment, our prioritization framework (Section 4) reveals that the highest-value action—XAI integration—is also most cost-effective. This finding challenges the assumption that effective cybersecurity AI requires substantial investment, demonstrating that meaningful improvements are accessible to resource-constrained organizations.

### 6.2 Limitations

**Secondary Analysis.** Our gap analysis and prioritization are based on secondary analysis of Ankalaki et al. (2025) rather than primary empirical research. While this approach efficiently leverages comprehensive survey findings, it inherits limitations of the underlying studies. Performance metrics reported may reflect optimistic experimental conditions rather than operational deployments.

**Generalizability Constraints.** The prioritization framework (Section 4) applies general scoring across dimensions. Individual organizations may have context-specific priorities that warrant different weightings. For example, healthcare organizations facing HIPAA compliance may prioritize privacy/ethics higher than our framework suggests, while financial institutions targeting APTs may elevate adversarial robustness.

**Rapidly Evolving Landscape.** Cyber threats and AI techniques evolve continuously. Our analysis, grounded in research through 2025, will become outdated as new attack vectors emerge and novel AI methods develop. The temporal obsolescence this study critique in datasets applies equally to analysis frameworks requiring periodic reassessment.

**Implementation Validation.** Our implementation roadmap (Section 5) provides theoretical guidance based on literature synthesis. Practical validation across diverse organizational contexts is needed to confirm effectiveness and identify context-specific adaptations required.

### 6.3 Future Research Directions

**Empirical Validation Studies.** Future research should empirically test the gap prioritization framework and implementation roadmap across organizations of varying sizes and sectors. Case studies documenting real-world deployments would validate our theoretical recommendations and identify practical challenges we may have overlooked.

**Automated Dataset Currency Assessment.** The temporal obsolescence problem (Gap 1) suggests need for automated mechanisms to assess dataset currency. Future work could develop algorithms that compare dataset attack patterns against current threat intelligence feeds, generating currency scores that guide dataset selection.

**Lightweight Adversarial Robustness.** The high cost of adversarial robustness testing (Section 4.2) limits adoption. Research should explore efficient robustness evaluation techniques suitable for resource-constrained organizations, potentially leveraging transfer learning or meta-learning to reduce computational requirements.

**Ethics-by-Design Frameworks.** While this study provide an ethical deployment checklist (Section 5.5), more comprehensive frameworks are needed that integrate ethical considerations throughout the ML lifecycle—from dataset collection through model deployment and monitoring. Research should develop concrete techniques for implementing principles identified by Ankalaki et al. (2025, Section X.D) in operational contexts.

**Hybrid Human-AI Decision Systems.** The interpretability challenge (Gap 3) points toward need for research on optimal human-AI collaboration in security operations. Future work should investigate how to design systems that leverage AI detection capabilities while preserving human judgment for complex decisions, particularly examining the role of XAI in supporting effective collaboration.

**Federated Learning for Privacy-Preserving Collaboration.** The privacy/ethics gap (Gap 5) and dataset obsolescence gap (Gap 1) could be simultaneously addressed through federated learning approaches that enable organizations to collaboratively train models on current threat data without sharing sensitive information. Research should explore federated learning architectures specifically designed for cyber attack prediction.

## 7. Conclusion

Artificial intelligence and machine learning have demonstrated remarkable capabilities for cyber attack prediction in research settings. The comprehensive survey by Ankalaki et al. (2025) documents impressive detection accuracies, with Random Forest achieving 99.92% on UKM-IDS20 and CNNs reaching 98.42% on CTU-13. However, a critical gap persists between these research results and practical deployment in operational security environments.

This paper has identified and analyzed five critical gaps hindering real-world implementation: temporal dataset obsolescence, narrow attack scope coverage, real-time interpretability challenges, inadequate adversarial robustness, and unaddressed privacy and ethical considerations. Through our gap prioritization framework, this study determined that XAI integration represents the highest-value action due to its combination of high impact, low cost, and moderate implementation time, while dataset obsolescence poses the most fundamental long-term threat to system effectiveness.

Our practical implementation roadmap translates comprehensive survey findings into actionable guidance. The dataset selection framework classifies 45 benchmark datasets by production readiness, enabling practitioners to make informed choices. Model selection guidelines match ML/DL techniques to organizational resources and priorities. The phased XAI integration strategy provides step-by-step implementation, while the ethical deployment checklist ensures responsible AI practices.

The path forward requires collaboration between researchers and practitioners. Researchers should prioritize developing production-ready solutions that address identified gaps—particularly creating current, comprehensive datasets and computationally efficient robustness testing methods. Practitioners should systematically address gaps using the prioritization framework, beginning with high-value, cost-effective improvements like XAI integration before investing in resource-intensive efforts like adversarial training.

Ultimately, effective cyber defense requires not just accurate attack prediction but interpretable, robust, ethical systems that security analysts trust and adversaries cannot easily evade. By bridging the research-practice gap, this study can translate the impressive capabilities demonstrated in research literature into operational security improvements that protect organizations against evolving cyber threats.

**Acknowledgements**

None

**References**

Ankalaki, S., Atmakuri, A. R., Pallavi, M., Hukkeri, G. S., Jan, T., & Naik, G. R. (2025). Cyber attack prediction: From traditional machine learning to generative artificial intelligence. *IEEE Access*, *13*, 44662-44706. https://doi.org/10.1109/ACCESS.2025.3547433

Butler, J. (2022). Cryptocurrency and blockchain technology in financial crime. *Journal of Financial Crime*, *29*(2), 639-654.

Cohen, L. E., & Felson, M. (1979). Social change and crime rate trends: A routine activity approach. *American Sociological Review*, *44*(4), 588-608. https://doi.org/10.2307/2094589

Cornish, D. B. (1994). The procedural analysis of offending and its relevance for situational prevention. *Crime Prevention Studies*, *3*, 151-196.

Cross, C., & Gillett, A. (2020). Exploiting trust for financial gain: A typology of cyberfraud. *Victims & Offenders*, *15*(1), 83-99. https://doi.org/10.1080/15564886.2019.1679280

Ferrag, M. A., Maglaras, L., Moschoyiannis, S., & Janicke, H. (2020). Deep learning for cyber security intrusion detection: Approaches, datasets, and comparative study. *Journal of Information Security and Applications*, *50*, Article 102419. https://doi.org/10.1016/j.jisa.2019.102419

Foley, S., Karlsen, J. R., & Putniņš, T. J. (2019). Sex, drugs, and bitcoin: How much illegal activity is financed through cryptocurrencies? *The Review of Financial Studies*, *32*(5), 1798-1853. https://doi.org/10.1093/rfs/hhz015

Leukfeldt, E. R., Lavorgna, A., & Kleemans, E. R. (2017). Organised cybercrime or cybercrime that is organised? An assessment of the conceptualisation of financial cybercrime as organised crime. *European Journal on Criminal Policy and Research*, *23*(3), 287-300. https://doi.org/10.1007/s10610-017-9338-7

Lusthaus, J., Varese, F., & Lavorgna, A. (2023). The structure and organization of cybercrime. *Annual Review of Criminology*, *6*, 409-427. https://doi.org/10.1146/annurev-criminol-030421-040120

Nguyen, T. T., & Luong, T. D. (2021). Cybercrime and cybersecurity in the digital age. *Computer Networks*, *195*, Article 108190. https://doi.org/10.1016/j.comnet.2021.108190

Okpa, J. T., Ajuonu, P., & Iheonu, C. (2022). Business email compromise: A systematic review of attack patterns and defence mechanisms. *Computers & Security*, *118*, Article 102732. https://doi.org/10.1016/j.cose.2022.102732

Wall, D. S. (2021). *Cybercrime and society* (3rd ed.). Sage Publications. (Cited in Ankalaki et al., 2025)

Wasserman, S., & Faust, K. (1994). *Social network analysis: Methods and applications*. Cambridge University Press. https://doi.org/10.1017/CBO9780511815478